\documentclass[
    amsmath,
    amssymb,
    reprint
]{revtex4-2}

\usepackage{graphicx}
\usepackage{dcolumn}
\usepackage{bm}

\usepackage[normalem]{ulem}

\usepackage{amsmath}
\usepackage{amssymb}
\usepackage{amsthm}
\usepackage{url}
\usepackage{xcolor}
\usepackage{hyperref}
\usepackage{mathrsfs}
\usepackage{multirow}
\usepackage{orcidlink}

\newcommand{\pars}[1]{\left( #1 \right)}
\newcommand{\bracs}[1]{\left[#1\right]}

\newcommand{\unit}[1]{\,{\rm{#1}}}
\newcommand{\sn}[1]{\times10^{#1}}
\renewcommand{\dfrac}[2]{\frac{d #1}{d #2}}

\newcommand{\parfrac}[2]{\pars{\frac{#1}{#2}}}
\newcommand{\Mc}{\mathcal{M}}

\renewcommand{\phi}{\varphi}

\newcommand{\Mwd}{M_{\rm WD}}
\newcommand{\dMwd}{\dot{M}_{\rm WD}}
\newcommand{\dMns}{\dot{M}_{\rm C}}
\newcommand{\Mns}{M_{\rm C}}
\newcommand{\Rwd}{R_{\rm WD}}
\newcommand{\dRwd}{\dot{R}_{\rm WD}}

\begin{document}

\preprint{APS/123-QED}

\title{Gravitational Wave Modeling of White-Dwarf--Compact-Object Binaries and Observational Outlook}

\author{Tristan S. Weaver$^{1,\,2}$\orcidlink{0009-0008-3529-6223}}\email{Tristan\_Weaver@psu.edu}%
\author{David Radice$^{1,\,2,\,3}$ \orcidlink{0000-0001-6982-1008}}%
\author{Donghui Jeong$^{1,\,2,\,4}$ \orcidlink{0000-0002-8434-979X}}
\author{Victor Liu$^{1}$ \orcidlink{0000-0003-4434-3921}}
\affiliation{$^{1}$Department of Astronomy and Astrophysics, The Pennsylvania State University, University Park, PA 16802, USA}
\affiliation{$^{2}$Institute for Gravitation and the Cosmos, The Pennsylvania State University, University Park, PA 16802, USA}
\affiliation{$^{3}$Department of Physics, The Pennsylvania State University, University Park, PA 16802, USA}
\affiliation{$^{4}$Korea Institute for Advanced Study, 85 Hoegi-ro, Dongdaemun-gu, Seoul 02455, Republic of Korea}

\date{\today}

\begin{abstract}
Binaries consisting of a white dwarf and a compact object (neutron stars, black holes, or sub-solar mass compact objects) display unique orbital evolution and gravitational wave signals.
We investigate the end stages of these systems by developing a novel integrator implementing mass transfer and 2.5-order Post-Newtonian kinematics. White-dwarf--compact-object binaries are subject to two phases: a slow inspiral during which binary evolution is dominated by weak-gravity radiation reaction and a quicker ``outspiral" dominated by mass transfer. These systems have unique gravitational waveforms with several characteristic features including an accumulation of signal near the maximum cutoff frequency set by the transition between in- and outspiral. We investigate trends in gravitational wave parameters based on component masses. In particular, maximum cutoff frequency depends strongly on white dwarf mass and compactness. It is unlikely that extragalactic binaries will be measured by LISA or current or future terrestrial detectors. However, white-dwarf--compact-object binary end stages will be an important source for future decihertz detectors such as DECIGO and LGWA: The final inspiral of tens of white-dwarf--neutron-star binaries will be detected annually by LGWA, and millions of inspirals along with thousands of outspirals will be detected by DECIGO. The ultimate fates of these binaries are diverse and dependent upon the intricacies of angular momentum loss. In particular, we show that white dwarfs with partners of sufficiently low/high mass can end their lives in 
mergers, while white dwarfs do not enter contact with partners of intermediate mass. The highest frequencies produced by white-dwarf--compact-object binaries in our tests are 10--20 Hz---incapable of mimicking sub-solar-mass compact object signals in terrestrial detectors or explaining S251112cm-like signals.

\end{abstract}

\maketitle


\section{Introduction} \label{sec:intro}
Since the first direct detection of a gravitational wave (GW150914; \cite{abbot+16}), the Laser Interferometer Gravitational-wave Observatory (LIGO; \cite{LIGO-15}) has observed a wide variety of compact binary mergers. By the end of its O3 run \cite{abbot+23}, LIGO had announced 90 candidate events, including 69 black-hole--black-hole (BH-BH), 4 black-hole--neutron-star (BH-NS), and 2 neutron-star--neutron-star (NS-NS) mergers confirmed. The cumulative number of event candidates through O4 is expected to be $>350$ 
\cite{adhicary+26,LVK_26a,LVK_26b}. These observations have been used to inform stellar mass BH and NS population models and binary synthesis models \cite{biscoveneau+23}, including the discovery of an apparent surplus of BHs with mass $\approx35\,M_\odot$ \cite{abbot+23,abbot+23b}. Observations of binaries containing neutron stars remain relatively rare \cite{salafia_25,fishbach+26} because gravitational-wave searches for low-mass objects are limited by survey volume (e.g., \cite{mandel+22}) with GW170817 remaining the only confirmed multi-messenger gravitational wave source \cite{abbot+17}. Together, LIGO, the Virgo interferometer 
\cite{Acernese+15}
 and the Kamioka Gravitational Wave Detector (KAGRA; 
\cite{KAGRA+19}) constitute the current network of terrestrial gravitational wave observatories (LVK; e.g., \cite{LVK_22}) and are sensitive in the frequency band $\sim10$--$1000\unit{Hz}$.     

LVK and planned terrestrial detectors such as the Einstein Telescope (ET; 
\cite{punturo+10}) and Cosmic Explorer (CE; \cite{evans+23}) are best suited to detect high frequency stellar mass compact object mergers. Detectors designed to probe different frequency ranges will be sensitive to different gravitational wave sources. The only non-LVK gravitational wave detection to date is the very low-frequency stochastic gravitational wave background \cite{agazie+23} from unresolved supermassive BH binaries by the North American Nanohertz Observatory for Gravitational Waves (NANOGrav; \cite{mcLaughlin_13}) and International Pulsar Timing Array (IPTA; 
\cite{manchester+13, perera+19}). 

The planned Laser Interferometer Space Antenna (LISA; 
\cite{amaro+17}) will probe frequencies between IPTA and LVK in the $10^{-4}$--$10^{-1}\unit{Hz}$ band. At these frequencies, LISA is expected to detect novel sources including massive BH binaries \cite{sesana+05, klein+16_long} and Extreme Mass Ratio Inspirals (EMRIs; \cite{amaro+07}). 
Furthermore, LISA is expected to detect Galactic binaries 
\cite{nelemans+01, kupfer+18, burdge+19, finch+23, kupfer+24} 
and their unresolved contribution to the millihertz stochastic gravitational wave background \cite{farmer+03}. These observations will constitute the first gravitational wave detections of white dwarfs (WDs). However, such systems evolve over timespans much longer than LISA's mission lifetime and will appear as stationary sources.

WDs are also predicted to emit gravitational waves in quickly evolving Type Ia supernovae systems, though these signals are expected to fall in the gap between LISA and terrestrial detector peak sensitivity \cite{falta+11,korol+24_SNe}. Observations of Type Ia supernovae will likely require a detector such as the proposed Deci-hertz Interferometer Gravitational wave Observatory (DECIGO; 
\cite{kawamura+21}) or the Lunar Gravitational Wave Antenna (LGWA; 
\cite{ajith+25}) sensitive in this gap. 

In this paper, we investigate WDs in binaries as dynamic gravitational wave sources; in particular, we study binaries containing a WD and a compact object. The same binaries that will be detected by LISA as stationary sources of gravitational waves harden under the influence of radiation reaction. Frequency evolves as $\dot f\propto f^{11/3}$ \cite{peters_64, blanchet+14,poisson+14}, so binaries of sufficiently high frequency evolve quickly enough to be measured in a detector lifetime. In this way, we can study WDs in close binary systems as analogues to the rapidly evolving compact object binaries detected by LVK.    

In addition to their status as potential novel gravitational wave sources, WDs are of interest in light of two recent sub-threshold sub-solar-chirp-mass event candidates reported by LVK \cite{LVK_25a, LVK_25b}. These events are low-significance and are likely of terrestrial origin with False Alarm Rates of $\approx5\unit{month}^{-1}$ and $6\unit{yr}^{-1}$, respectively, far below typical significance of reported events (cf., \cite{abbot+19}). However, if they were of astrophysical origin, one explanation for such events would be the presence of sub-solar mass (SSM) compact objects (e.g., \cite{nitz+21,abbot+22, LVK_23,miller_24,prunier+24, riajul+26, viera+26}). WDs are the densest known objects besides BHs and NSs, and their mass distribution peaks in the sub-solar range ($\approx 0.6\,M_\odot$; \cite{kepler+07,tremblay+16}). Thus, they are a natural alternative to SSM compact objects in explaining signals with reported sub-solar chirp mass. 

To model gravitational waveforms for a compact object binary in the final strong-gravity regime, it is necessary to solve Einstein's Equations through methods of numerical relativity (e.g., \cite{lehner_01,baumgarte+10,lehner+14}). Such methods are computationally expensive, and solving the entire inspiral phase of binary evolution with numerical relativity would be prohibitive. Instead, at wider separations under the influence of weaker gravity, it is sufficient to use Post-Newtonian (PN) approximations to evolve the binary and calculate the waveform (e.g., \cite{blanchet+14}).

These approximations are generally applied under the point mass assumption for compact objects. However, a WD will undergo significant tidal interaction and mass loss at sufficiently close separation. In this regime, PN kinematic equations alone are insufficient to describe system dynamics. Instead, WD evolution is modeled as a self-gravitating fluid under the influence of PN radiation reaction. Recent examples include \cite{chen_25b} who model the electromagnetic signal due to tidal interactions in WD-NS binaries, \cite{kremer+17,sala+25} who model the gravitational waveform in WD-WD interactions, and \cite{paschalidis+09,tauris_18,yu+26} who model gravitational waveforms in WD-NS binaries. These studies use forms for the PN equations derived under the assumption of Keplerian orbits in place of fully general PN equations (see \cite{poisson+14}). Additionally, some employ expensive full fluid-mechanical methods to solve the problem of WD 
density distribution \cite{chen_25b,bobrick+17} (see also \cite{paschalidis+09, dan+12, pakmor+12, sato+16} on WD binaries).       

In this paper, we present a model to calculate gravitational waveforms for circular WD-compact object binaries undergoing mass transfer with the goal of characterizing waveforms and forecasting observation feasibility in current and future detectors. In \S\ref{sec:methods}, we introduce our code---a fast, fluid mechanics-informed, semi-analytic model incorporating mass transfer physics (\S\ref{sec:methods-MT}), and general 2.5PN radiation reaction kinematics (\S\ref{sec:methods-RR})---and we detail our prescriptions for numerical integration and waveform calculation (\S\ref{sec:methods-GWs}). In \S\ref{sec:results}, we discuss system evolution (\S\ref{sec:results-evolution}) and gravitational wave characteristics (\S\ref{sec:results-gw-char}) and trends (\S\ref{sec:results-gw-trends}), the observational consequences of which we discuss in \S\ref{sec:discussion}. We conclude in \S\ref{sec:conclusion}. Throughout this paper, we will refer to ``compact objects" (COs) as NSs, BHs, and hypothetical non-WD sub-solar mass objects (i.e., we distinguish between compact objects which we treat as point particles and WDs which we treat as extended bodies).         

\section{Methods} \label{sec:methods}
We evolve a circular WD-compact object orbit considering the effects of mass transfer and gravitational wave radiation reaction. We assume isotropic emission such that the orbit remains circular. In the following sections, we introduce the semi-analytic integration methods used to evolve the system and to calculate resultant gravitational waves.

\subsection{Mass Transfer}\label{sec:methods-MT}
We first model the effect of mass transfer on a system with WD of mass and radius $\Mwd,\,\Rwd$, respectively, compact object of mass $\Mns$, and orbital separation of $a$. For a binary with mass ratio $q:=\frac{M_{\rm WD}}{\Mns}$, the Eggleton approximation for the Roche Limit \cite{eggleton_83} is given as:
\begin{equation}
    R_L=a\frac{0.49 q^{2/3}}{0.6q^{2/3}+\ln\pars{1+q^{1/3}}}\label{eqn:RL}
\end{equation}
The instantaneous Keplerian period of the system is:
\begin{equation}
    P\pars{\Mwd,\,\Mns,a}=\frac{2\pi a^{3/2}}{\sqrt{G\pars{\Mwd+\Mns}}} \label{eqn:period}
\end{equation}
Then, we approximate the mass loss of the WD as:
\begin{equation}
    \dMwd=-\frac{A}{P}\int_{R_L}^{\Rwd}4\pi r^2\,\rho_{\rm WD}(r)\,dr \label{eqn:dMwd}
\end{equation}
$A$ in Equation~\ref{eqn:dMwd} is a dimensionless mass transfer rate constant of order 10 (e.g., \cite{ritter_88, kolb+90, ryu+25}). We take an $n=1.5$ polytropic density profile (e.g., \cite{hamada+61}; see \S\ref{sec:app-PI} for further discussion) to calculate the mass contained in the shell outside of the Roche Lobe, $\int_{R_L}^{\Rwd}4\pi r^2\,\rho_{\rm WD}(r)\,dr$. 

The equilibrium radius of a non-spinning WD is given as a function of $\Mwd$ by \cite{nauenberg_72}:
\begin{align}
    R_{\rm WD,\,0}=7792\unit{km}&\parfrac{M_{\rm WD}}{0.7\,M_\odot}^{-1/3}\nonumber\\
    &\cdot\sqrt{1-\parfrac{M_{\rm WD}}{M_{\rm Ch}}^{4/3}}\parfrac{\mu_e}{2}^{-5/3}\label{eqn:Rwd0}
\end{align}
with chemical composition of a carbon-oxygen WD assumed ($\mu_e=2$) and Chandrasekhar mass $M_{\rm Ch}=1.44\,M_\odot$ \cite{chandrasekhar_31}. Choosing instead the mass-radius relation of \cite{verbunt+88} has no significant effect on the system. We evolve the WD's radius toward the equilibrium radius over a dynamical timescale, $\tau_{\rm dyn}:=\sqrt{\frac{R_{\rm WD}^3}{GM_{\rm WD}}}$:
\begin{equation}
    \dRwd=\frac{R_{\rm WD,\,0}-R_{\rm WD}}{\tau_{\rm dyn}} \label{eqn:dRwd}
\end{equation}

We assume conservative mass transfer onto the compact object until the Eddington accretion limit \cite{eddington_26,frank+85} is reached:
\begin{equation}
    \dMns=\min\pars{-\dMwd,\,\dot M_{\rm edd,\,C}} \label{eqn:dMns}
\end{equation}
with:
\begin{equation}
    \dot M_{\rm edd,\,C}=\frac{4\pi G m_{ p}}{\sigma_T\eta_{\rm acc}c}\Mns\label{eqn:eddington}
\end{equation}
Respectively, $m_{ p}$ and $\sigma_T$ are proton mass and Thomson cross-section, and we take $\eta_{\rm acc}=0.1$ (e.g., \cite{inogamov+99,sibgatullin+00,chaskina+17}). In practice, we find that $|\dMwd|\gg\dot M_{\rm edd,\,C}$ shortly after initial Roche Lobe Overflow (RLOF). Then, Equation~\ref{eqn:dMns} reduces to $\dMns=\dot M_{\rm edd,\,C}$. The choice of $\eta_{\rm acc}$ to within an order of magnitude has no significant effect on the system.

Finally, we model the evolution of the binary separation due to mass loss. In general, mass, angular momentum, and separation evolution in a binary are related by \cite{soberman+97}:

\begin{equation}
    2\,\frac{\dot J}{J}=\frac{\dot a}{a}+2\,\frac{\dot M_{\rm WD}}{M_{\rm WD}}+2\,\frac{\dot \Mns}{\Mns}-\frac{\dot M_{\rm WD}+\dot \Mns}{M_{\rm WD}+ \Mns}-\frac{2 e\dot e}{1-e^2}\label{eqn:general}
\end{equation}
$\beta\equiv \left|\frac{\dMns}{\dMwd}\right|=\min\pars{\frac{\dot M_{\rm edd,\,C}}{|\dot M_{\rm WD}|},1}$ parameterizes mass conservation in the system and $\gamma$ parameterizes the specific angular momentum of ejected material such that angular momentum loss due to non-conservative mass transfer is:
\begin{equation}
    \frac{\dot J_{\rm MT}}{J}=\gamma(1-\beta)\frac{\dot M_{\rm WD}}{M_{\rm WD}+\Mns}\label{gamma}
\end{equation}
Assuming isotropic mass loss such that eccentricity, $e=0$, is constant, Equation~\ref{eqn:general} simplifies to: \begin{widetext}
\begin{equation}
    \dot a_{\rm MT}=-2a\frac{\dot M_{\rm WD}}{M_{\rm WD}}\bracs{1-\beta\frac{M_{\rm WD}}{\Mns}-\pars{1-\beta}\pars{\gamma+\frac{1}{2}}\frac{M_{\rm WD}}{M_{\rm WD}+\Mns}}\label{eqn:da_MT}
\end{equation}
\end{widetext}
where $\dot a_{\rm MT}$ is the evolution of the binary separation due to effects associated with mass transfer. See \cite{pols_14} for a pedagogical introduction to binary evolution.

\subsubsection{Angular Momentum Loss Modes}\label{sec:methods-MT-L}
In general, it is necessary to model a system with fluid dynamics to determine precisely how angular momentum is lost \cite{soberman+97}. Under certain simplifying assumptions, however, it is possible to calculate analytically the angular momentum lost and to provide a simple closed-form solution for $\gamma$. Two such assumptions are Jeans Mode and Isotropic Re-emission.

In the case of Jeans Mode angular momentum loss, matter leaving the system is ejected isotropically in the rest frame of the donor (see \cite{tauris+06,saladino+18,schroder+21} for physical examples). Then, the expelled matter carries with it specific angular momentum equal to that of the WD, implying $\gamma=\frac{\Mns}{\Mwd}$. 

Isotropic Re-emission involves matter falling from the donor onto its partner and subsequently being expelled from the system \cite{soberman+97,king+95,misra+20}, carrying away specific angular momentum equal to that of the CO with $\gamma=\frac{\Mwd}{\Mns}$. 

These mechanisms are alternatively referred to as ``Fast" and ``Slow" Modes, respectively, in reference to the angular-momentum-loss rate when $q<1$. Between these two extremes lies  the Intermediate Mode in which mass is deposited into a circumbinary ring with radius $a_{\rm ring}$ \cite{vanWinckel_03,shao+012,buj+13}; in this configuration, $\gamma=\frac{\pars{\Mns+\Mwd}^2}{\Mwd\Mns}\sqrt{\frac{a_{\rm ring}}{a}}$.

It has been found that assumptions about angular momentum loss strongly affect the ultimate fates of binary star systems \cite{willcox+23,gallegos-garcia+24}; indeed, we find this to be the case in \S\ref{sec:discussion-L}. However, we also find these assumptions do not strongly affect other conclusions of our work such as general gravitational wave trends and observational consequences (\S\ref{sec:results},~\ref{sec:discussion-rates},~\ref{sec:discussion-SSM}). The results reported in these sections are from our Jeans Mode calculations.

\subsection{Radiation Reaction}\label{sec:methods-RR}
Angular momentum loss due to the emission of gravitational waves hardens the binary and is dependent upon system kinematics. To fully determine position, we introduce azimuthal angle $\theta$ which evolves as:
\begin{equation}
    \dot \theta=\frac{2\pi}{P}\label{eqn:dt}
\end{equation}
under the adiabatic assumption ($\frac{\dot a}{a}\ll\dot \theta$, e.g., \cite{peters_64,dosopoulou+16}).

We calculate the quadrupole moment, $Q$, and traceless quadrupole moment, $I$, for point mass WD and CO:
\begin{align}
    Q^{ij}&:=\int\,d^3\vec{x}\,\rho(\vec{x})\,x^ix^j\nonumber\\
    &=\mu\, x^i(a,\theta)\,x^j(a,\theta)\nonumber\\
    I^{ij}&:=Q^{ij}-\frac{1}{3}\delta^{ij}Q^k_{\,\,k}\label{eqn:quadrupole}
\end{align}
with reduced mass, $\mu:=\frac{\Mwd\Mns}{\Mwd+\Mns}$. For rectangular coordinates on the orbital plane, we can choose:
\begin{equation}
    \begin{cases}
        x=a\cos\theta\\
        y=a\sin\theta\\
        z=0
    \end{cases} \label{eqn:cartesian}
\end{equation}

To leading (2.5 Post-Newtonian) order, the angular momentum loss is determined by the second and third derivatives of the quadrupole moment \cite{peters_64}:  
\begin{equation}
    \dot{J}_{RR} = \frac{-2G}{5c^5} \left( \ddot{I}^{xp} \dddot{I}^y_{\;\;p} - \ddot{I}^{yp} \dddot{I}^x_{\;\;p} \right)+\mathscr{O}(c^{-7}) \label{eqn:dJ_RR}
\end{equation}
where second and third derivatives of $I$ are calculated by applying \texttt{ForwardDiff.jl} \cite{revels+16} to Equation~\ref{eqn:quadrupole} in conjunction with Equations~\ref{eqn:dMwd}--\ref{eqn:dt}. 

Finally, combining Equations~\ref{eqn:general} and~\ref{eqn:dJ_RR} yields orbital hardening due to radiation reaction to order $c^{-5}$:
\begin{equation}
      \dot a_{\rm RR}=\frac{-4aG}{5Jc^5}(\ddot{I}^{xp}\dddot I^y_{\,\,\,p}-\ddot{I}^{yp}\dddot I^x_{\,\,\,p})\label{eqn:da_RR}
\end{equation}
Combining Equations~\ref{eqn:da_MT} and~\ref{eqn:da_RR}, the total rate of change of the binary separation is:
\begin{widetext}
  \begin{align}
    \dot a&=\dot a_{\rm MT}+\dot a_{\rm RR}\nonumber\\&=-a\left\{2\frac{\dot M_{\rm WD}}{M_{\rm WD}}\left[1-\beta\frac{M_{\rm WD}}{\Mns}-(1-\beta)\left(\gamma+\frac{1}{2}\right)\frac{M_{\rm WD}}{M_{\rm WD}+\Mns}\right]+\frac{4G}{5Jc^5}(\ddot{I}^{xp}\dddot I^y_{\,\,\,p}-\ddot{I}^{yp}\dddot I^x_{\,\,\,p})\right\}\label{eqn:da}
\end{align}  
\end{widetext}

\subsection{Integration and Gravitational Waveforms}\label{sec:methods-GWs}
Given a set of initial conditions, $\pars{\Mwd,\,\Mns}$, we begin our integration at a separation, $a_{\rm init}$, corresponding to initial RLOF: $R_L(a_{\rm init};\Mwd,\,\Mns)=R_{\rm WD,\,0}(\Mwd)$. 

\begin{figure*}
    \centering
    \includegraphics[width=0.95\linewidth]{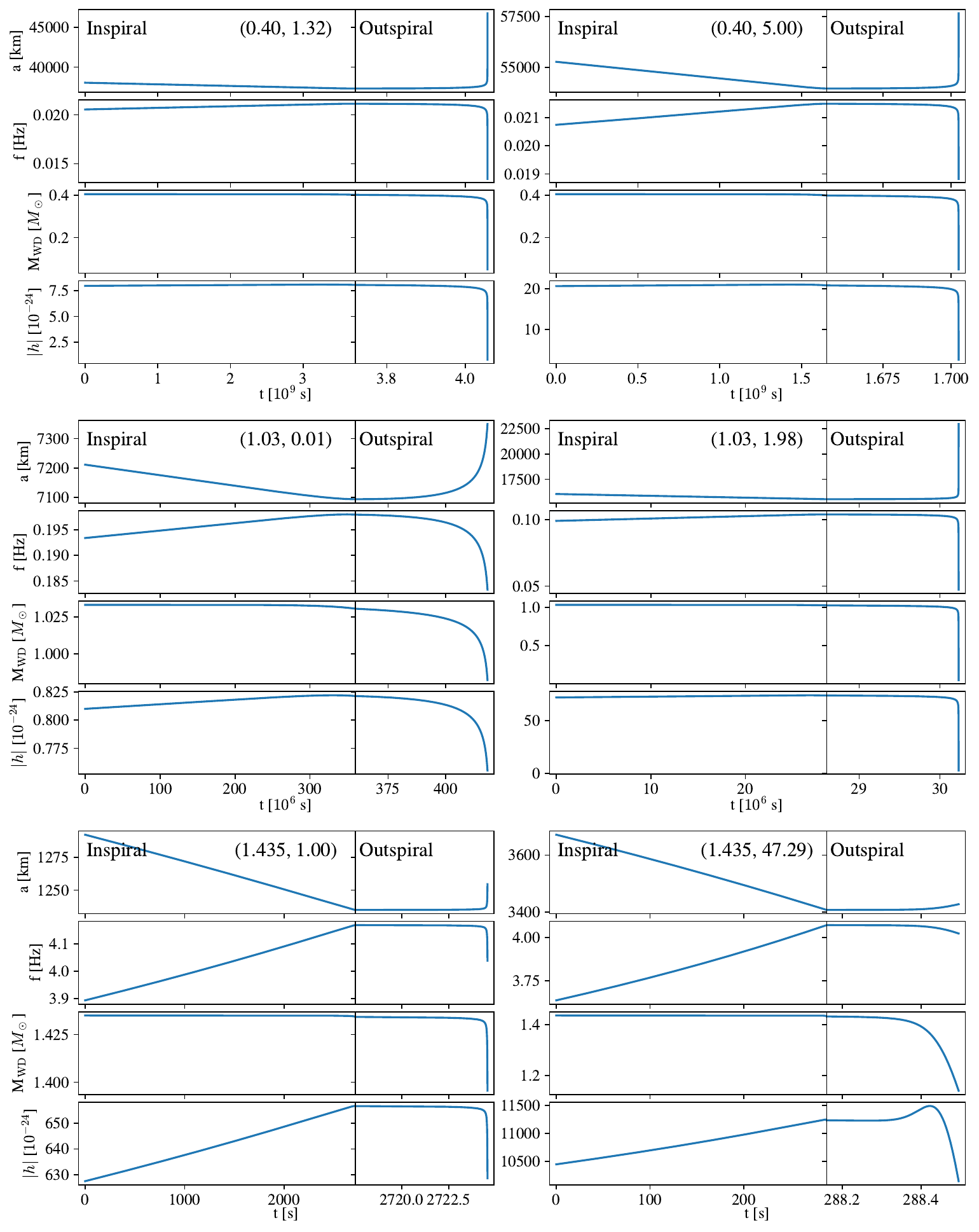}
    \caption{Evolutionary tracks for six representative binary systems labeled as $(\Mwd,\Mns)$. Binary separation, $a$, gravitational wave frequency, $f$, WD mass, $\Mwd$, and strain amplitude, $|h|$, for a face-on binary at $D_L=1\unit{Mpc}$ are plotted against time since RLOF, $t$, from top to bottom in each set of panels.  Each panel is divided into inspiral (left) and outspiral (right) portions. Note that the timescale changes between in-/outspiral portions. Corresponding $h_c$ tracks are shown in Fig.~\ref{fig:Fig_3}.}
    \label{fig:Fig_1}
\end{figure*}

We evolve our system with the \texttt{DifferentialEquations.jl} \cite{rackauckas+17} implementation of \cite{verner_10}. Given the very long time for some of these systems to evolve (see Fig.~\ref{fig:Fig_1}), it is impractical to evolve the system purely on period-scale timesteps. When the system is evolving slowly, we integrate with respect to separation, $a$, by transforming the evolution equations as $\dfrac{}{a}=\pars{\dfrac{a}{t}}^{-1}\dfrac{}{t}$ (e.g., \cite{Messina+19,sayeb+21}). Our integration pipeline follows the procedure:
\begin{enumerate}
    \item $a$-basis integration from initiation of mass transfer to just before separation minimum (Fig.~\ref{fig:Fig_1}),
    \item $t$-basis integration from just before to just after separation minimum,
    \item $a$-basis integration from just after separation minimum to when timesteps become sub-period ($\Delta t<P$), and 
    \item $t$-basis integration from when $\Delta t<P$ to end (\S\ref{sec:app-stopping}). 
\end{enumerate}

In steps 1--3, we do not evolve $R_{\rm WD}$ dynamically as in Equation~\ref{eqn:dRwd} for purposes of numerical stability; instead, we take $R_{\rm WD}=R_{\rm WD,\,0}$ as in Equation~\ref{eqn:Rwd0}. As $\Delta t\gg \tau_{\rm dyn}$ in these regimes, it is reasonable to instantaneously evolve $R_{\rm WD}$ to its ``natural" state.

We continue integrating until $\Mwd<0.05\,M_\odot$ or $\Rwd\geq a$ (\S\ref{sec:app-stopping}). Because the lightest WD observed has $\Mwd=0.17\,M_\odot$ \cite{kilic+07}, any models for $\Mwd\lesssim0.17\,M_\odot$ contain inherent uncertainty (e.g., \cite{althaus+13} does not include models below $0.15\,M_\odot$). However, we find that the general conclusions drawn in \S\ref{sec:results} and~\ref{sec:discussion} do not depend strongly on the assumed minimum WD mass below $\Mwd=0.17\,M_\odot$.

Our model calculates $\ddot I^{ij}$ during integration in order to derive a timeseries gravitational wave for an observer with known luminosity distance, $D_L$, and orientation relative to the binary, $(\iota,\,\phi)$, via the quadrupole approximation \cite{einstein_18,blanchet+14}:
\begin{equation}
     h^{ij}_{\rm (TT)}(t)=\frac{2G}{c^4D_L}\ddot I^{ij}_{\rm (TT)}\pars{t-D_L/c}\label{eqn:quadrupole_formula}
\end{equation}

\begin{table*}[]
    \centering
    \begin{tabular}{c c c c c }
         Category & Type& Masses Sampled $[M_\odot]$\\\hline
         \multirow{2}{*}{$\Mwd$} & Low-to-High Mass & \{0.30, 0.40, 0.51, 0.61, 0.72, 0.82, 0.93, 1.03, 1.14, 1.24, 1.35\}\\
 & Ultra-relativistic & \{1.41, 1.43, 1.435, 1.4375, 1.439\}\\\hline
 \multirow{3}{*}{$\Mns$} & SSM & \{0.01, 0.1, 0.5\}\\
  & NS & \{1.0,1.33,1.65,1.98,2.3\}\\
  & BH & \{5.0, 10.57, 22.36, 47.29, 100.0\}\\\hline
    \end{tabular}
    \caption{Initial masses sampled. \label{tab:mass_grid}}
\end{table*}

Assuming that the detector is aligned with either the plus or cross polarization of the gravitational wave, characteristic strain is \cite{moore+15}:
\begin{equation}
    h_{c}(f)=2f\,\tilde h(f)=2f\left|\int\,dt \,h_{+/\times}(t)W(t)\,e^{-2ift}\right|\label{eqn:hc}
\end{equation}
where $W(t)$ is a windowing function (e.g., Hann; see \cite{allen+12, talbot+21}). In practice, given that evolution is near-adiabatic, we calculate the Fourier transform using the Stationary Phase Approximation on monotonic segments of the waveform \cite{sathya+91,cutler+94,Droz+1999}:
\begin{equation}
    \tilde{h}(f)=\left|h(t(f))\right|\sqrt{\frac{2\pi}{|\dot f|}}\,e^{i\bracs{\theta(t(f))-2\pi ft(f)+\phi_0}}\label{eqn:SPA}
\end{equation}
where $\left|h(t(f))\right|$ is the amplitude of the gravitational wave strain, $h(t)$, when the instantaneous frequency of the gravitational wave is $f$, and $\phi_0$ is a global phase at the beginning of the monotonic segment.

\section{Results}
\label{sec:results}
Using the integration pipeline described in the previous section, we evolve systems over a grid of 16 initial values for $\Mwd$ and 13 initial values for $\Mns$. Our test set of WDs includes uniformly sampled low-to-high-mass WDs and five ultra-relativistic WD masses. Our test set of compact objects includes three hypothetical sub-solar mass (SSM) objects, five uniformly sampled neutron star masses, and five stellar black hole masses sampled logarithmically. With our integrator, it is also possible for us to extend our work to include intermediate-mass black holes $\pars{{\rm IMBHs;\,}10^2\,M_\odot<M_{\rm IMBH}<10^5\,M_\odot}$ which we will pursue in future work. A full list of initial masses sampled is given in Table~\ref{tab:mass_grid}. We assume angular momentum loss through 
Jeans Mode (see \S\ref{sec:discussion-L}) and mass transfer coefficient $A=10$ (see \S\ref{sec:app-MT}).

In this section, we present the evolution of WD-CO systems (\S\ref{sec:results-evolution}) and characteristic features of their gravitational waveforms (\S\ref{sec:results-gw-char}). We conclude by examining trends in resultant gravitational waveforms as a function of system initial conditions (\S\ref{sec:results-gw-trends}).

\subsection{Sample System Evolution}
\label{sec:results-evolution}
In Figure~\ref{fig:Fig_1}, we plot evolutionary tracks for six representative binary systems ranging from low-mass to relativistic WDs over a wide range of mass ratios. Each track exhibits the characteristic behavior of these systems: a slow inspiral during which radiation reaction dominates the system's kinematics, followed by a quicker period of binary widening (``outspiral") dominated by mass transfer, ending in the WD's decay. 

For systems with high-mass WDs, the outspiral phase can end early due to an expanding WD engulfing its companion. System fates are further discussed in \S\ref{sec:discussion-L-merger}.

These systems range over a wide variety of timescales from  minutes to centuries. The time from initial RLOF to turnaround strongly depends on $\Mwd$ while $\Mns$ has an approximately order-of-magnitude effect. 

\subsection{Gravitational Waveform Characteristics}

\label{sec:results-gw-char}

\begin{figure}
    \centering
    \includegraphics[width=\linewidth]{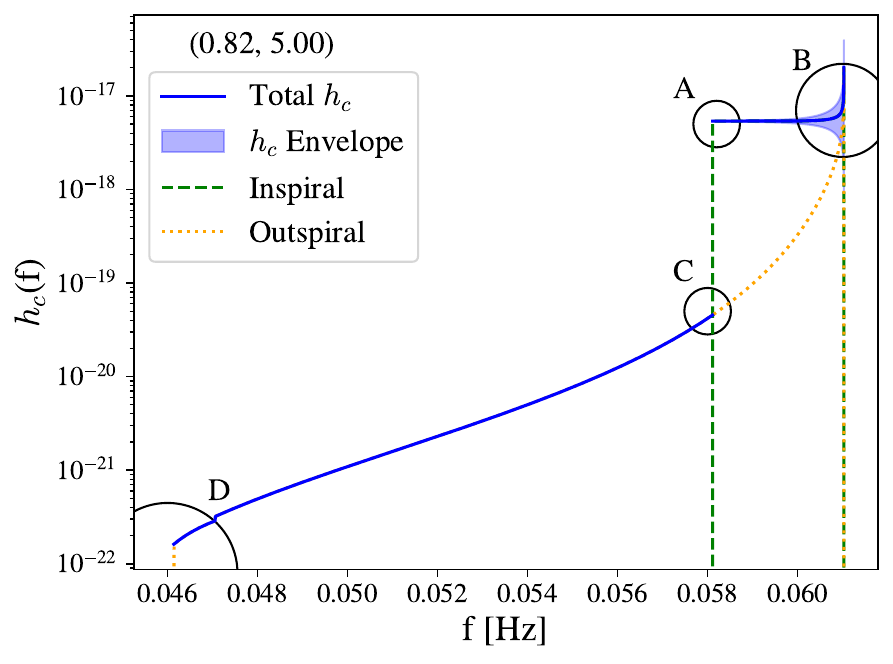}
    \caption{Sample characteristic strain curve for a system with initial $(\Mwd,\Mns)=(0.82,5.00)\,M_\odot$ observed face-on at a distance $D_L=1\unit{Mpc}$ for a four-year duration. Salient features are marked in the plot chronologically from beginning to end: A) Low-frequency cutoff due to finite observation time; B) $f_{\rm max}$ cutoff due to turnaround and $h_c$ envelope due to inspiral/outspiral interference; C) Outspiral $h_c$ contribution dominates; and D) the integration stops at $\Mwd=0.05\,M_\odot$.}
    \label{fig:Fig_2}
\end{figure}

WD-CO events demonstrate several unique, characteristic features displayed in  Figure~\ref{fig:Fig_2}. In line with the nominal four-year mission lifetime of LISA  
\cite{amaro+17}, all $h_c$ waveforms presented in this paper are calculated over a four-year evolution period. If the outspiral takes longer than two years, the waveform is calculated centered on $t(f_{\rm max})$. Otherwise, the waveform is calculated for the four years leading up to WD dissipation.

Due to this finite observation time, the inspiral phase has a low-frequency cutoff corresponding to the start of the observation (Feature A in Fig.~\ref{fig:Fig_2}). The system evolves along the $h_c\propto f^{-1/6}$ track characteristic of point mass inspirals \cite{maggiore_07,sathya+09} until mass transfer becomes comparable to radiation reaction in its influence on system kinematics. 

As mass transfer becomes more important, $\dot a\rightarrow 0$ and frequency evolves slowly compared to evolution due to the effects of radiation reaction alone. This leads to a buildup in $h_c$ at the turnaround frequency, $f_{\rm max}$, up to $\approx 50$ times higher than expected due to $h_c\propto f^{-1/6}$. $f_{\rm max}$ also serves as a characteristic high frequency cutoff (Feature B in Fig.~\ref{fig:Fig_2}).   

Also shown in Feature B is an envelope for values of $h_c$ of the total event where the inspiral and outspiral contribute comparably to $h_c$. Considering the contributions of the inspiral and outspiral to $\tilde h(f)$ separately, we can write $\tilde h_{\rm In/Out}(f)=|\tilde h_{\rm In/Out}(f)| e^{i\phi_{\rm In/Out}}$. Because $\dfrac{\ln f}{t}\ll P^{-1}$ during the inspiral, the phase, $\phi_{\rm In/Out}(f)$, varies very quickly as a function of $f$ near $f_{\rm max}$ (Eqn.~\ref{eqn:SPA}). Therefore, $\phi_{\rm In}(f)$ and $\phi_{\rm Out}(f)$ are uncorrelated, and:
\begin{align}
     |\tilde h(f)|&=\left||\tilde h_{\rm In}(f)|e^{i\phi_{\rm In}}+|\tilde h_{\rm Out}(f)|e^{i\phi_{\rm Out}}\right|\nonumber\\
     &\in\left[|\tilde h_{\rm In}(f)|-|\tilde h_{\rm Out}(f)|,|\tilde h_{\rm In}(f)|+|\tilde h_{\rm Out}(f)|\right]\label{eqn:hc_env}
\end{align}
with $|\tilde h(f)|$ oscillating rapidly within this envelope. As a result, integrated quantities of total $h_c$ (e.g., SNR) are determined by the greater of the two components: $\int\,df\,g(f)h_c({\rm f)}=\int\,df\,g(f)\max \{h_{c,{\rm In}}(f),h_{c,{\rm Out}}(f)\}$ for arbitrary function $g$. In practice, it is easier to consider $h_{c,\,{\rm In/Out}}$ for the in-/outspiral as separate events.

During the outspiral phase, the binary widens quickly compared to inspiral hardening, meaning that $h_c$ from the outspiral is suppressed by orders of magnitude compared to the inspiral. In addition, as the outspiral continues, binary widening accelerates, so $h_c$ decreases as $f$ decreases. The outspiral may produce lower frequencies than the inspiral does during the observation window (Feature C in Fig.~\ref{fig:Fig_2}). 
Finally, the signal tails off as the WD decays until $\Mwd<0.05\,M_\odot$.

\begin{figure*}
    \centering
    \includegraphics[width=0.75\linewidth]{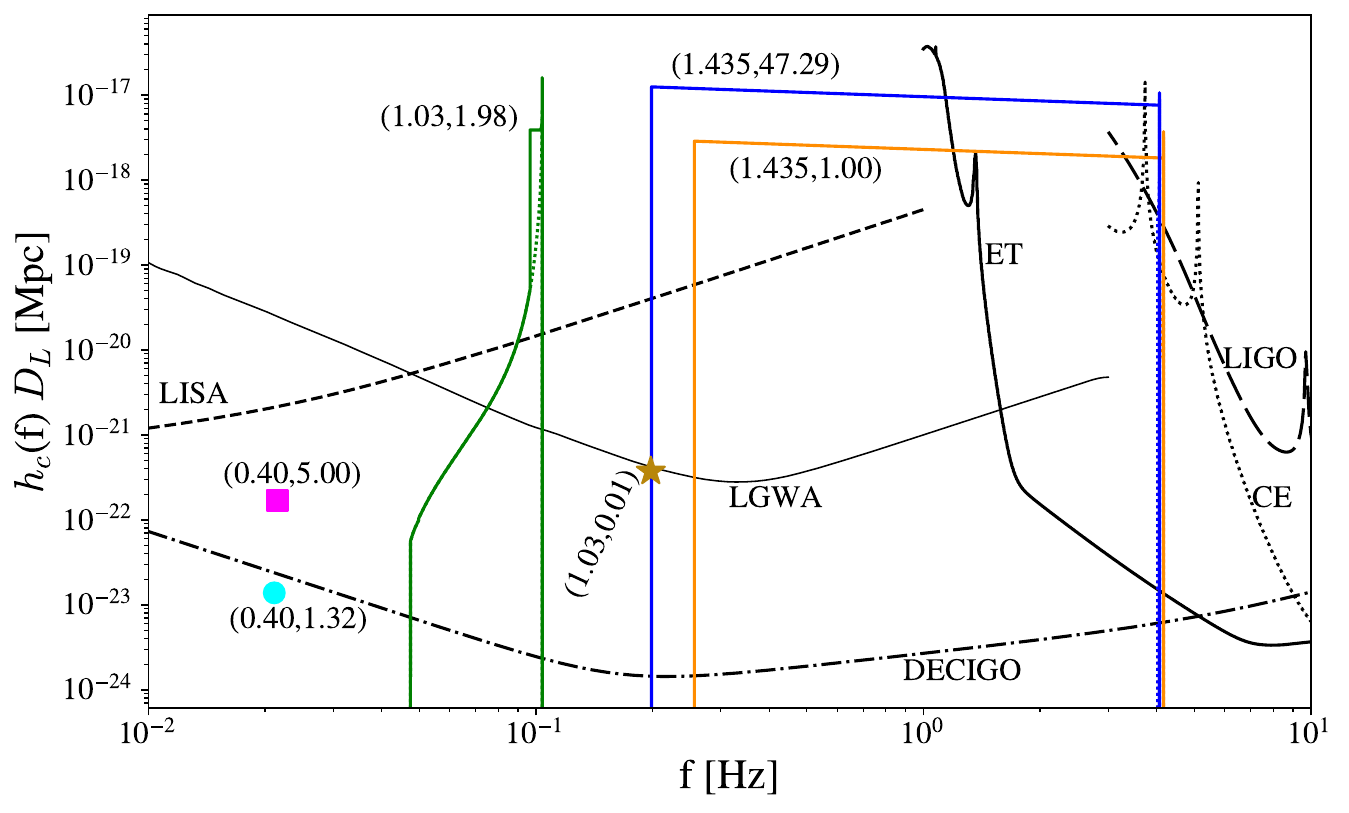}
    \caption{Characteristic strain tracks for six representative binary systems shown in Figure~\ref{fig:Fig_1} assuming face-on observation at a distance $D_L=1\unit{Mpc}$ for a four-year duration. The three least massive binaries are effectively stationary over the four-year observation window, and their integrated strain ($\int h_c\,df$) is plotted at their $f_{\rm max}$. Outspiral $h_c$ tracks are denoted with a dotted line.}
    \label{fig:Fig_3}
\end{figure*}

Figure~\ref{fig:Fig_3} shows the characteristic strain curves of the representative $(\Mwd,\Mns)$ binaries of Figure~\ref{fig:Fig_1} face-on at a distance of $D_{L}=1\unit{Mpc}$. The sensitivity of LIGO's O4 is retrieved from \texttt{PyCBC} \cite{barsotti+18,biwer+19}; the planned sensitivities of LISA, ET-D, and CE (40 km) are retrieved from \texttt{kuibit} 
\cite{robson+19,colpi+24,hild+11,reitze+19, bozzola_21}; the planned sensitivity of LGWA-Si is retrieved from \texttt{GWFish} 
\cite{ajith+25, dupletsa+23}; and DECIGO's sensitivity is analytically modeled as:

\begin{widetext}
    \begin{equation}
   S_n=7.05\sn{-48}\bracs{1+\parfrac{f}{f_p}^2}+\frac{4.80\sn{-51}\unit{Hz}^4}{f^4} \bracs{1+\parfrac{f}{f_p}^2}+\frac{5.33\sn{-52}\unit{Hz}^4}{f^4}
\end{equation}
\end{widetext}
with transfer frequency, $f_p=7.36\unit{Hz}$, over a range $f\in[10^{-3},10^2]\,\unit{Hz}$ \cite{yagi+11}.

The timespan from the beginning of mass transfer to the end of the binary varies by seven orders of magnitude between the evolutionary tracks sampled in Figure~\ref{fig:Fig_1}. As a result, while systems with the most massive WDs see a significant change in frequency, the lightest systems experience virtually no evolution during the four-year observation window. For the three lightest systems, we plot $\pars{f_{\rm max},\int\,df\,h_c(f)}$ rather than an $h_c$ track which is effectively a delta spike. Each of these persistent source systems is in the LISA band, but even at 1 Mpc under ideal observing conditions, they are too quiet for LISA detection. On the other hand, two out of three of these sources are detectable by DECIGO with the $(\Mwd,\Mns)=(1.03,0.01)\,M_\odot$ binary detectable as far as $D_L\sim30\unit{Mpc}$. A $(\Mwd,\Mns)=(1.03,0.01)\,M_\odot$ binary in M31 ($D_L=761\unit{kpc}$; \cite{li+21}) would be marginally detectable by LGWA.

The relativistic WD sources have $f_{\rm max}\approx 4\unit{Hz}$, potentially making them marginally detectable by ET. The four-year evolutionary track begins in the 
LGWA band, providing an early warning for contemporaneous terrestrial observatories (e.g., \cite{sesana_16,sesana_17,tso+19,klein+22}). It is possible that the entire event, from inspiral to the WD engulfing its companion, may be detected by DECIGO.  

In principle, these events are also at the proper frequency to be detected by LISA before subsequent end-stage detection in terrestrial detectors. However, LISA detection would require a local source ($D_L\sim 30\unit{Mpc}$). Given the low expected rate of these events in the local universe \cite{nelemans+01b}, this scenario is unlikely over LISA's lifetime (see \S\ref{sec:discussion-rates}).

\subsection{Gravitational Waveform Trends}

\begin{figure}
    \centering
    \includegraphics[width=\linewidth]{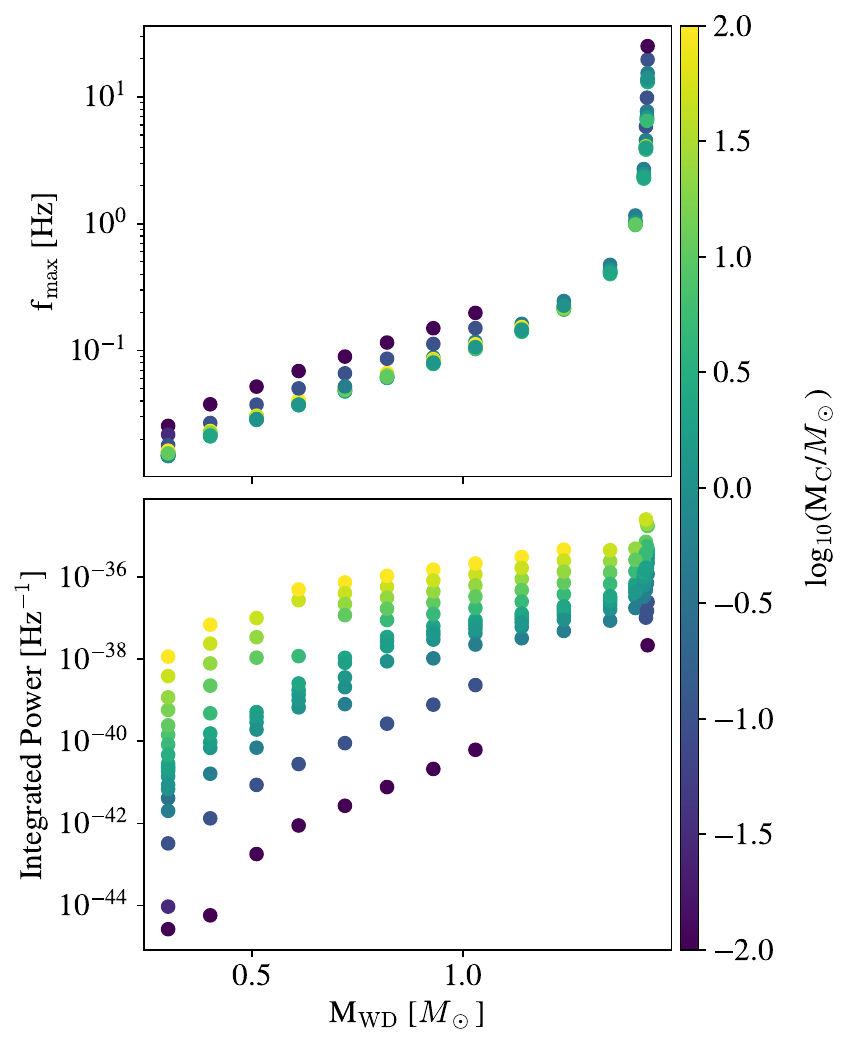}
    \caption{Cutoff frequency, $f_{\rm max}$, and integrated signal power as functions of WD and CO initial mass. }
    \label{fig:Fig_4}
\end{figure}

\begin{figure}
    \centering
    \includegraphics[width=\linewidth]{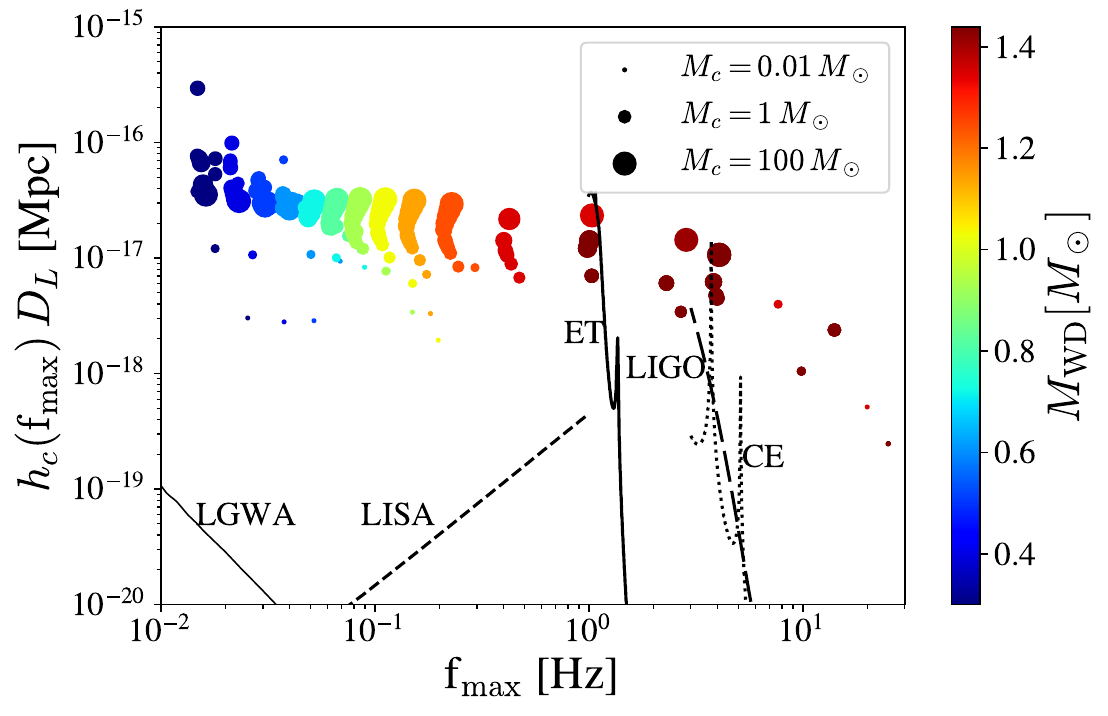}
    \caption{Location of the high frequency cutoff peak (Feature B in Fig.~\ref{fig:Fig_2}) superimposed on sensitivity curves for LISA and terrestrial detectors. Initial WD mass is coded in color and initial CO mass is coded in size.}
    \label{fig:Fig_5}
\end{figure}
\label{sec:results-gw-trends}
Figures~\ref{fig:Fig_4} and~\ref{fig:Fig_5} depict general trends in gravitational wave frequency and strength for our sampled WD-CO systems. Initial $\Mwd$ is the most important predictor of gravitational wave parameters. This is because $R_{\rm WD}\equiv R_{\rm WD}(\Mwd)$ for a WD in equilibrium (Eqn.~\ref{eqn:Rwd0}), which, in turn, determines the separation and frequency at which initial RLOF occurs. $f_{\rm max}$ is tightly controlled by initial $\Mwd$ and varies between $\sim10^{-2}$--$10\unit{Hz}$ from the least to the most massive WDs. $\Mns$ provides a comparably modest factor of $\lesssim 2$ variation in $f_{\rm max}$ for fixed $\Mwd$.  

The strength of the gravitational wave signal is still largely determined by $\Mwd$, but $\Mns$ is more important in this case than in determining $f_{\rm max}$. Integrated power over a four-year observation ($\int h_c^2\,df$; Fig.~\ref{fig:Fig_4}) is generally higher for high initial $\Mwd$ and $\Mns$. The effect of $\Mns$ is larger for low $\Mwd$ systems. 

Figure~\ref{fig:Fig_5} shows the value of $h_c$ at $f_{\rm max}$ (Feature B in Fig.~\ref{fig:Fig_2}) for each evolutionary track. We observe different trends in $h_c(f_{\rm max})$ for systems of low and high $\Mwd$: High-$\Mwd$ systems tend to have higher $h_c(f_{\rm max})$ when partnered with high-mass compact objects. In contrast, with the exception of systems with sub-solar-mass compact objects, low-$\Mwd$ systems tend to have higher $h_c(f_{\rm max})$ when partnered with low-mass compact objects. Figure~\ref{fig:Fig_5} also reveals that for systems with $0.8\,M_\odot\lesssim \Mwd\lesssim1.35\,M_\odot$, mass ratios of $1\lesssim q\lesssim3$ tend to minimize $f_{\rm max}$ compared to more extreme mass ratios.

\begin{figure}
    \centering
    \includegraphics[width=\linewidth]{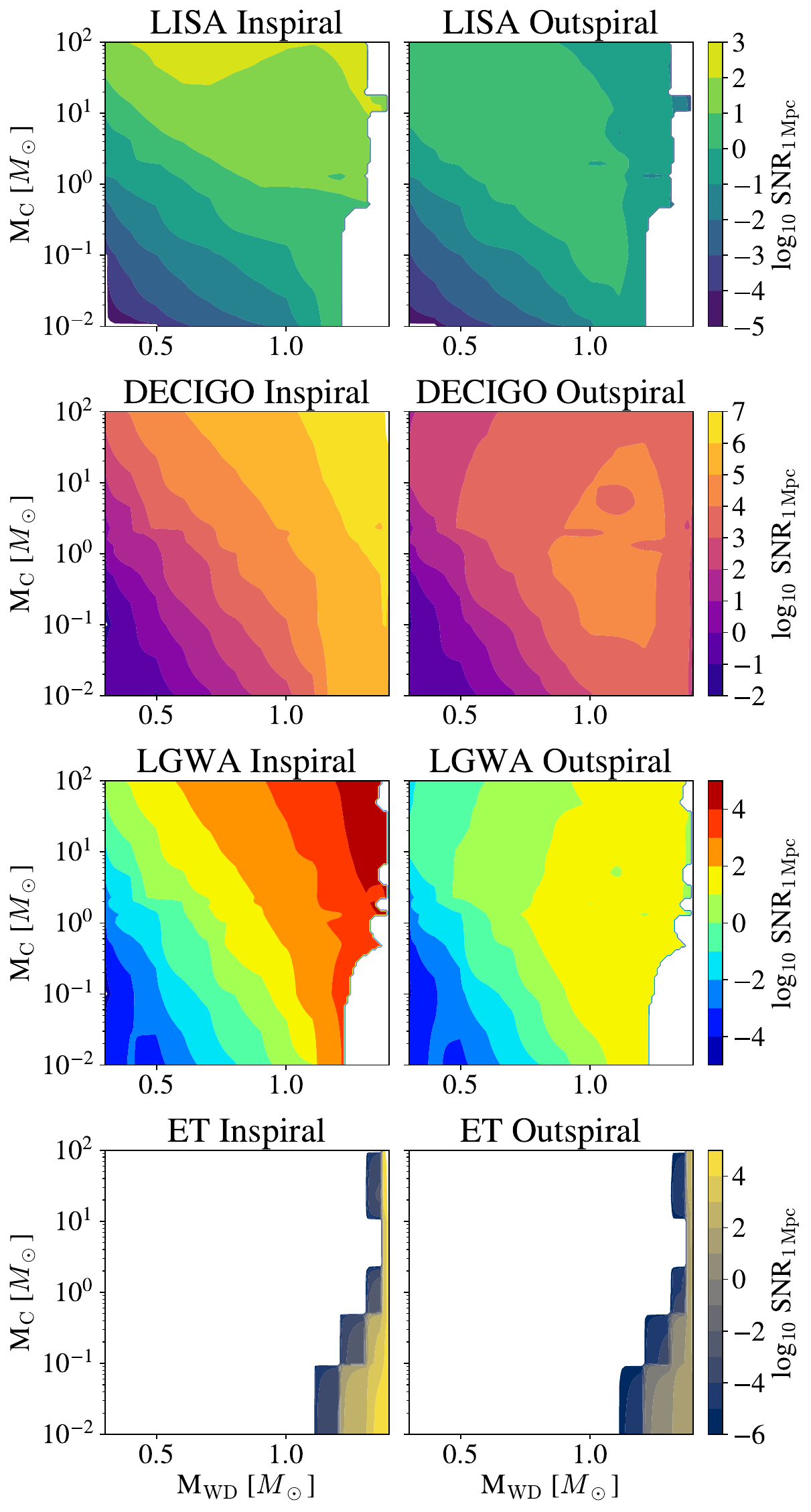}
    \caption{Signal-to-noise ratio (SNR) of initial $(\Mwd,\Mns)$ configurations in their inspiral (\textit{left}) and outspiral (\textit{right}) phases for LISA, DECIGO, LGWA, and ET. For LISA, DECIGO, and LGWA, a four-year observation with 75\% duty cycle is assumed. SNR is calculated for a face-on binary at $D_L=1\unit{Mpc}$. SNR is only calculated for a detector if the system's $f_{\rm max}$ falls within the band of the relevant detector, and SNR below the threshold indicated by respective colorbars is not shown.}
    \label{fig:Fig_6}
\end{figure}

\section{Discussion}\label{sec:discussion}
\subsection{Observed Event Rates}\label{sec:discussion-rates}

To better understand the prospects of detecting gravitational waves from WD-CO systems, to model their contribution to the unresolved stochastic gravitational wave background, and to inform future detector design, we make order-of-magnitude predictions for the detection rate of WD-CO binary end stages (inspiral turnarounds and outspiral WD disintegrations).

We estimate the volume rate of WD-NS/BH turnarounds to order-of-magnitude as:
\begin{equation}
    R=\frac{\nu_{\rm MW}}{M_{\rm MW}}\cdot\rho_{\rm crit,\,M}\pars{1-f_{\rm IGM}}
\end{equation}
where $\nu_{\rm MW}=1.4\sn{-4},\,1.9\sn{-6}\unit{yr}^{-1}$ is, respectively, the Galactic merger rate of NS-WD and BH-WD binaries predicted by \cite{nelemans+01b}, 
$M_{\rm MW}=10^{12}\,M_\odot$ is the mass of the Galaxy \cite{wang+20}, $\rho_{\rm crit,\,M}=2.66\sn{-30}\unit{g\,cm}^{-3}$ is the matter density of the local universe \cite{adame+25}, and $f_{\rm IGM}=0.76$ is the fraction of baryons in the intergalactic medium \cite{connor+25}. This implies a NS-WD end stage rate of order 
$R_{\rm NS-WD}\sim10^3\unit{Gpc}^{-3}\unit{yr}^{-1}$ and a BH-WD merger rate of order  $R_{\rm BH-WD}\sim20\unit{Gpc}^{-3}\unit{yr}^{-1}$.

In Figure~\ref{fig:Fig_6}, we calculate the face-on SNR for $(\Mwd,\Mns)$ configurations in LISA, DECIGO, LGWA, and ET at $D_L=1\unit{Mpc}$. We assume a detection threshold of SNR $\geq 8$ (e.g., \cite{allen+12, abbot+16}) and we apply LISA's projected duty cycle of $75\%$ \cite{amaro+22} to it, LGWA, and DECIGO. In this way, we can interpret the DECIGO inspiral panel to mean that DECIGO will be able to detect massive WD-CO binaries at cosmological distances, and we can read detection horizon distances off of the other seven panels as $r_{\rm horizon}=\frac{{\rm SNR}_{1\unit{Mpc}}}{8}\unit{Mpc}$. Averaging for geometrical effects, the effective search volume of an event is \cite{finn+93}:
\begin{equation}
    V_{\rm eff}\approx\frac{4\pi}{3}\parfrac{r_{\rm horizon}}{2.26}^3
\end{equation}

Informed by \cite{korol+2024}, we model the distribution of masses in WD-NS systems as a bivariate normal distribution: \begin{widetext}
    \begin{equation}
    P(\Mwd,M_{\rm NS})=\mathcal{N}(\Mwd,M_{\rm NS};\hat{M}_{\rm WD}=0.9,\sigma_{M_{\rm WD}}=0.25,\hat{M}_{\rm NS}=1.3,\sigma_{M_{\rm NS}}=0.25,\rho=0.3)\label{eqn:P_WD-NS}
\end{equation}
\end{widetext}
where $\hat{M}_{\rm WD/NS}$ signifies the mean mass of the respective populations, $\sigma_{M_{\rm WD/NS}}$ signifies the standard deviation in mass in the respective population, and $\rho$ signifies covariance between WD and NS mass.
Due to stellar evolution leading to WD-NS binaries, these WDs are biased toward higher masses with a peak at $\approx 0.9\,M_\odot$ compared to the overall WD population, which peaks at $\approx0.6\,M_\odot$ \cite{kepler+07,tremblay+16}. For WD-BH systems, we take $P(\Mwd,M_{\rm BH})=P(\Mwd)P(M_{\rm BH})$ given in \cite{shao+21}.

Finally, we calculate extragalactic detection event rates in Table~\ref{tab:rates} as:
\begin{align}
    R_{\rm detect}=R&\int\,d\Mwd\nonumber\\ & \int d\Mns\,V_{\rm eff}(\Mwd,\Mns)P(\Mwd,\Mns)\label{eqn:detection_rate}
\end{align}

WD-CO binaries are not powerful enough to be detected at distances farther than $\sim10$-$30\unit{Mpc}$ by LISA. Consequently, it is not expected that LISA will observe the end stages of a WD-CO binary over its observing run. The expected observation timescale for terrestrial detectors is greater than the age of the Universe. 

By contrast, WD-CO end stages will be important sources for decihertz detectors. Crucially for its science goals related to WD studies 
\cite{ajith+25,benetti+26}, LGWA will detect the final inspirals of tens of WD-NS binaries annually, comparable to LVK's O4 rate of detection for BH-BH mergers 
\cite{adhicary+26, LVK_26a}.

DECIGO will detect the final stages of WD-NS and WD-BH inspirals at cosmological distances. 
It will be necessary to develop advanced data analysis such as improved global fit and iterative subtraction (``peeling") techniques \cite{cornish+03,cutler+06,littenberg+23} in order to process the millions of overlapping signals that DECIGO will be sensitive to simultaneously. These efforts will be critical to ensure that these signals do not obscure subtle cosmological sources key to DECIGO's science case (e.g., inflationary gravitational waves 
\cite{seto+01,kawamura+11}). 
Outspirals will be individually detected at rates of 
$\sim2\sn{3}$ and $2\unit{yr}^{-1}$
for WD-NS and WD-BH binaries, respectively.

\begin{table}[]
    \centering
    \begin{tabular}{c c c c c}
         \multirow{2}{*}{Detector}
& \multicolumn{2}{c}{WD-NS} 
& \multicolumn{2}{c}{WD-BH}\\

& Inspiral
& Outspiral
& Inspiral
& Outspiral\\\hline
LISA & $8\sn{-6}$ & $6\sn{-8}$ & $10^{-8}$ & $6\sn{-11}$\\
DECIGO & $>10^6$ & $2\sn 3$ & $>10^6$ & 2\\
LGWA & 80 & $5\sn{-4}$ & $6\sn{-2}$ & $6\sn{-8}$ \\
ET/CE/LIGO & $\sim0$ & $\sim0$ & $\sim0$ & $\sim0$\\\hline
    \end{tabular}
    \caption{Order-of-magnitude extragalactic WD-CO end stage detection rates in $\unit{yr}^{-1}$. \label{tab:rates}}
\end{table}

\subsection{Effect of Angular Momentum Loss Mode}
\label{sec:discussion-L}
In the main section of this paper, we assume angular momentum loss via Jeans Mode. 
However, the angular momentum carried away by non-conservative mass transfer, and thus the appropriate form of $\gamma$, remain uncertain \cite{willcox+23}. Existing binary evolution studies employ a variety of angular-momentum-loss prescriptions (cf., $\gamma=1$ is fixed in \cite{yu+26}), while the true outflow geometry and torque likely require hydrodynamic simulations to model self-consistently \cite{scherbak+25}. Despite uncertainties, the treatment of angular momentum loss is critical in determining binary fate (e.g., \cite{soberman+97}).   

In this section, we probe the effect of this assumption by performing a set of tests with angular momentum loss driven by Isotropic Re-emission. In particular, we examine how the fates of these systems differ depending on angular-momentum-loss mode assumption. In \S~\ref{sec:discussion-L-merger}, we discuss initial conditions under which the system can be driven toward a merger. In \S~\ref{sec:discussion-L-UCXB}, we describe a behavior in the Isotropic Re-emission case that is not observed under Jeans Mode: A phase of stable mass transfer resulting in an ultracompact X-ray binary (UCXB).    

\begin{figure*}
    \centering
    \includegraphics[width=0.85\linewidth]{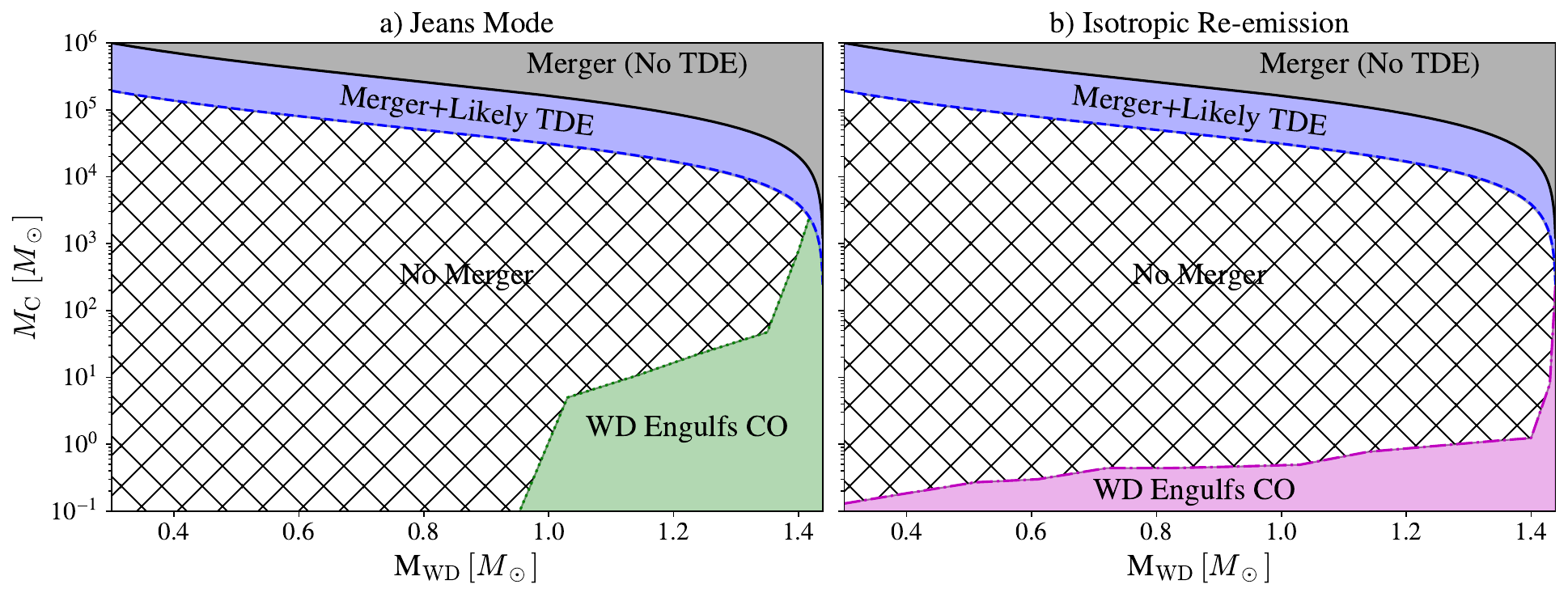}
    \caption{Fate of WD-CO binaries by initial mass under the assumption of Jeans Mode (\textit{left}) and Isotropic Re-emission (\textit{right}) angular-momentum-loss modes. WDs in circular orbits around IMBHs ($\Mns\gtrsim10^5\,M_\odot$) are forced into mergers due to gravitational plunge within ISCO radius. Under Jeans Mode angular momentum loss, high-mass WDs can be driven to engulf NS and stellar mass BH partners. Under Isotropic Re-emission, high-mass WDs engulf low-mass NSs, and near-Chandrasekhar WDs can additionally engulf BHs. Systems with COs too heavy for WD engulfment and too light for a strong gravity-forced merger (the parameter space labeled ``No Merger") may end their lives as (mass-transfer-driven) TDEs, UCXBs, or with the WD stripped to a brown-dwarf-like remnant supported by ideal gas pressure (see \S\ref{sec:app-stopping}). Further refinement of the ``No Merger" parameter space is left for subsequent studies.}
    \label{fig:Fig_8}
\end{figure*}

\subsubsection{Mergers and  WD Engulfments}
\label{sec:discussion-L-merger}

A merger detected by LVK is characterized by an inspiral due to radiation reaction followed by a final plunge with separation approaching zero under the effects of strong gravity (e.g., \cite{buonanno+99,buonanno+00,ori+00}). There are two classes of events that we may consider analogous to mergers in the case of WD-CO binaries. In the first such class, the WD is drawn  beyond the 
innermost stable circular orbit (ISCO) of a BH partner under the effects of radiation reaction, and a large fraction of the WD is accreted onto the BH. In the second such class, a WD loses mass and expands until it envelops its partner. 

In the first class of merger-like events, WDs enter the strong gravity regime of their BH partner and are either tidally disrupted or experience an LVK-like merger with $a\rightarrow0$. As BH mass increases, tidal forces on the WD near the BH event horizon decrease. For IMBHs sufficiently massive, a WD on a circular orbit will pass inside of the ISCO radius \cite{misner+73}:
\begin{equation}
    R_{\rm ISCO}=3R_{ S}=\frac{6GM_{\rm IMBH}}{c^2}\label{eqn:ISCO}
\end{equation}
before overflowing its Roche Lobe (Eqn.~\ref{eqn:RL}). A WD on a circular orbit with $a<R_{\rm ISCO}$ will undergo a final rapid plunge onto the BH. Significant mass transfer may occur between $R_{\rm ISCO}$ and $R_S$, resulting in shredding and a tidal disruption event (TDE; \cite{rosswog+09}). At even higher IMBH masses, a WD can cross the Schwarzschild radius, $R_S$, before mass transfer is initiated, resulting in a ``dark" merger without any electromagnetic counterpart (e.g., \cite{luminet+89,kesden+12, macLeod+14,kawana+18,maguire+20}). 

We assume non-spinning Schwarzschild BHs to calculate the ``Merger + TDE" and ``Dark Merger (No TDE)" regions in Figure~\ref{fig:Fig_8}; assumed angular-momentum-loss mode does not affect these regions. Dynamics are substantially more complicated in situations with realistic spinning Kerr BHs, particularly for generic EMRIs due to frame dragging and multi-frequency orbital motion \cite{amaro+07, katz+21}.

Conditions for the second class of events (WD engulfment) are strongly dependent on the angular-momentum-loss mechanism. In the case of Jeans Mode (Fig.~\ref{fig:Fig_8}a), light WDs ($\Mwd\lesssim 0.9\,M_\odot$) initiate mass transfer at sufficiently wide separations such that they cannot expand enough to undergo WD engulfment. Heavy WDs ($\Mwd\gtrsim 1.0\,M_\odot$) will engulf NS partners and  some stellar BH partners. Due to the lower separation between them, higher mass WDs can engulf higher mass BHs. 
Systems with initial $\Mwd\gtrsim1.42\,M_\odot$ must either pass inside of ISCO or undergo WD engulfment.

WD engulfment conditions change significantly under the assumption of Isotropic Re-emission (Fig.~\ref{fig:Fig_8}b). For mass ratio $q>\frac{1+\sqrt{17}}{4}\approx1.28$, mass transfer has the effect of accelerating binary hardening ($\dot a_{\rm MT}<0$). Therefore, for any $\Mwd$, there is a (potentially sub-solar) $\Mns$ with which the WD can engulf. However, for $q<1$, Isotropic Re-emission results in a faster increase in binary separation at a given $\dfrac{\ln \Mwd}{t}$ than does Jeans Mode, suppressing the highest $\Mns$ at which mergers can occur for heavy WDs ($\Mwd\gtrsim1.0\,M_\odot$). This effect is also what allows UCXB formation under Isotropic Re-emission and not under Jeans Mode (\S\ref{sec:discussion-L-UCXB}). WD masses of $\Mwd>1.25\,M_\odot$ are required to engulf NS partners, and masses of $\Mwd\gtrsim1.42\,M_\odot$ are required to engulf BHs.

\subsubsection{UCXB}
\label{sec:discussion-L-UCXB}
\begin{figure}
    \centering
    \includegraphics[width=\linewidth]{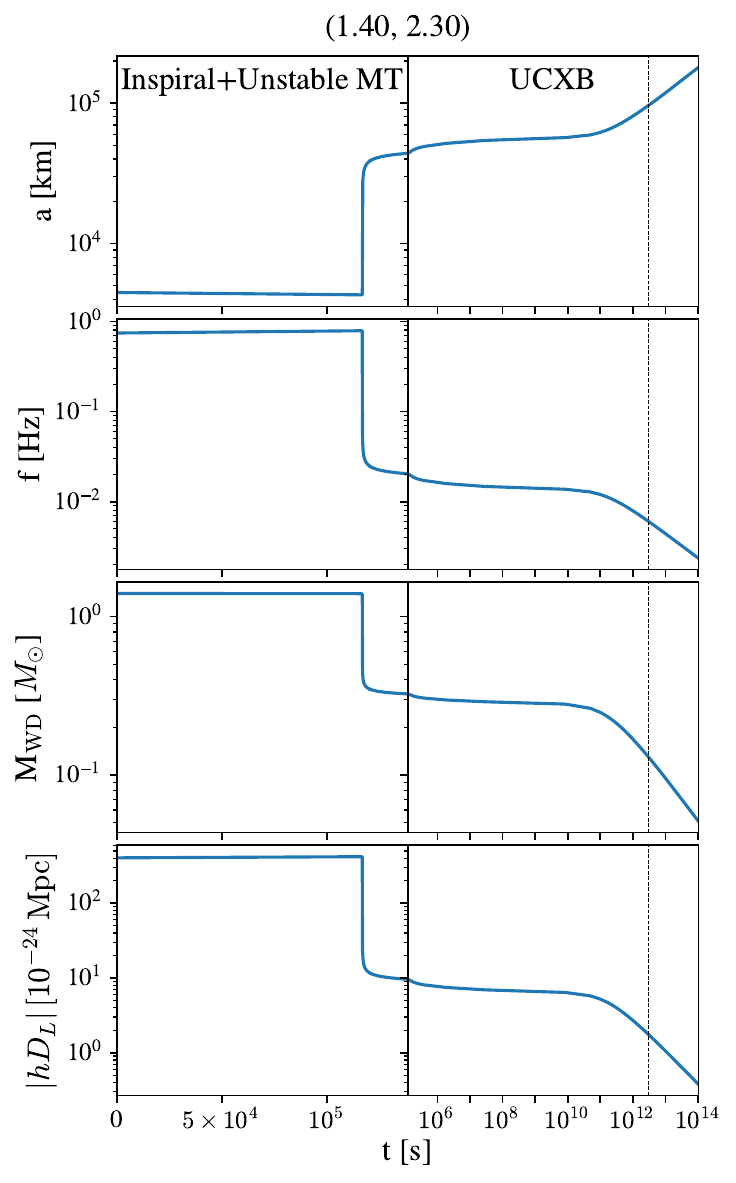}
    \caption{Evolution curve analogous to Fig.~\ref{fig:Fig_1} for a system with initial $(\Mwd,\,\Mns)=(1.40,\,2.30)\,M_\odot$ evolved under the Isotropic Re-emission assumption. Unlike the Fig.~\ref{fig:Fig_1} Jeans Mode examples, this system undergoes two stages with distinct behavior: Phase I which encompasses the final stage of inspiral, stall, and rapid widening coupled with unstable mass transfer; and Phase II which is characterized by a comparatively long, slow phase of stable mass transfer and orbital widening (i.e., a UCXB). Note the difference in scale between the time axis in the left and right panels. The system has LISA horizon $r_{\rm horizon}\gtrsim20\unit{kpc}$ for $t\lesssim3\sn{12}\unit{s}$ denoted by the vertical dashed line. At $t\gtrsim10^{13}\unit{s}$, $|\dot M_{\rm WD}|\lesssim \dot M_{\rm Edd.,\,NS}$, and mass transfer becomes conservative.}
    \label{fig:Fig_9}
\end{figure}
In Figure~\ref{fig:Fig_9}, we show the evolution after initial Roche Lobe overflow for a system with masses $(\Mwd,\,\Mns)=(1.40,\,2.30)\,M_\odot$ and Isotropic Re-emission angular momentum loss. As in the representative cases shown in Figure~\ref{fig:Fig_1}, this system experiences initial hardening due to radiation reaction and then undergoes a short period of rapid mass loss.

The system in Figure~\ref{fig:Fig_9} then reaches a steady state at $\Mwd\approx0.35$--$0.39\,M_\odot$. The 
binary separation is such that mass transfer and radiation reaction are nearly balanced. The system widens and loses mass very slowly, and mass transfer becomes quasi-stable; this stage constitutes a UCXB.  
This UCXB stage lasts orders of magnitude longer than 
the longest-lived systems from Figure~\ref{fig:Fig_1}.
The system in Figure~\ref{fig:Fig_9} would be detectable across the Galaxy with LISA horizon distance $r_{\rm horizon}>20/10/2\unit{kpc}$ for $t\lesssim100/300/3000\unit{kyr}$. 

Stable UCXB formation requires that mass transfer is stabilized through Isotropic Re-emission from the CO; in the case where transferred matter is instead lost via Jeans Mode, the system does not survive to form a long-term stable UCXB \cite{soberman+97, vanHaaften+12, vanHaften+12b, bobrick+17}. Consistent with these previous studies, we do not find any UCXBs formed under the Jeans Mode assumption.

Of the initial conditions tested with Isotropic Re-emission, the system in Figure~\ref{fig:Fig_9} is not the only configuration that forms a UCXB, though UCXB formation is not universal across the initial conditions we tested. In future work, more detailed analysis can find the parameter space in which UCXBs are formed (e.g., Fig.~\ref{fig:Fig_8}).
   
A quasi-stable close orbit adiabatically widening for $\Mwd\lesssim0.35\,M_\odot$ is consistent with models of UCXBs under the assumption of Isotropic Re-emission \cite{yungelson+02,vanHaaften+12}. However, fluid simulations including the contributions of accretion disks and impact physics on angular momentum loss suggest an upper bound near $\Mwd\lesssim0.2\,M_\odot$ \cite{bobrick+17}. There is active debate regarding the threshold for stability with \cite{chen+22} predicting 
UCXBs with WD masses up to $0.45\,M_\odot$. 

The standard evolutionary path of a UCXB involves a low-mass He WD entering Roche Lobe contact and then forming a stable system in which radiation reaction and mass transfer are in steady state \cite{deloye+03}. The period of intense, unstable mass transfer widening the orbit before UCXB formation as in our scenario is absent (e.g., \cite{paschalidis+09}), but \cite{bobrick+17} and \cite{wang+21} suggest an analogous stage for more massive, semi-degenerate He stars. 
Additionally, the UCXB implied by Figure~\ref{fig:Fig_9} would be unique in its donor's carbon-oxygen/ONeMg composition, though not singularly: carbon-oxygen donors have been observed (e.g., 4U 0614+091, 4U 1543-624; \cite{werner+06,nelemans+06}), and there have even been ONeMg candidates (e.g., 4U 1626-67; \cite{schulz+01}), raising additional questions regarding UCXB formation pathways \cite{nelemans+10}.

The system described in Figure~\ref{fig:Fig_9} with gravitational wave frequency $f\sim 10\unit{mHz}$ and orbital period $T\sim200\unit{s}$ is significantly faster than the shortest observed period UCXB (4U 1820-30) with frequency $f=2.92\unit{mHz}$ and period $T=685\unit{s}$ \cite{stella+87}. Because the high-frequency phase of a UCXB's life is shorter than the low-frequency phase, observations are biased against high-frequency systems; a system can start its life with a higher frequency than 4U 1820-30, but the system is more likely to be observed later in its long-lived, low-frequency state \cite{yang+26}.

The comparison of Figure~\ref{fig:Fig_9} to the literature highlights the importance of $\gamma$ in determining WD-CO end stage behavior. UCXBs are expected to be important Galactic gravitational wave sources in the LISA band with EM counterpart lifetimes of $\sim10\unit{Myr}$ and gravitational wave detectability up to $\sim100\unit{Myr}$ \cite{nelemans+04,vanHaaften+12,vanHaaften+13, qin+24}.
The systems usually modeled as LISA sources in these analyses, however, are wider and lower-frequency than the system in Figure~\ref{fig:Fig_9}, which reaches $f\sim10\unit{mHz}$ near the high-frequency end of the expected Galactic UCXB population \cite{stella+87,chen+20, qin+24}. The evolutionary pathways leading to ultracompact X-ray binaries remain uncertain, with competing formation channels (e.g., white dwarf, helium star, evolved main sequence, dynamical systems) and differing assumptions regarding mass-transfer stability leading to substantially different predictions for the observable UCXB population \cite{pod+02,deloye+03,vanHaaften+12,vanHaaften+13,bobrick+17,chen+22,wang+26,yang+26}.

\subsection{WD-CO Systems as Impostor SSM Signals}
\label{sec:discussion-SSM}

In the last year, two candidate sub-threshold, sub-solar-mass events have been reported. The first, S250818k, was reported with a coincident electromagnetic transient \cite{LVK_25a}. However, \cite{gillanders+25} and \cite{yang+25} concluded that the transient was an unrelated supernova rather than a kilonova counterpart to S250818k. S251112cm was later observed without any coincident electromagnetic transient \cite{LVK_25b, LSST_25}. Compared to S250818k, it has a higher detection significance and, if of astrophysical origin, a greater likelihood of involving a binary with a sub-solar-mass component (False Alarm Rate $\approx0.16\unit{yr}^{-1},\,P_{\rm sub-solar}>0.99$). Theoretical processes have been proposed for the origins of SSM COs, including primordial black holes \cite{carr+74,escriva+24}, boson stars \cite{liebling+12}, and exotic neutron stars \cite{chen+25,krnjaic+26}. However, there is no confirmed astrophysical process to produce SSM COs; indeed, the lightest electromagnetically confirmed CO is a NS with mass $\approx1.17\,M_\odot$ (J0453+1559's companion; \cite{martinez+15}).

LVK data from these two events have not yet been publicly released, and it seems unlikely that S251112cm will be found to have a true astrophysical source. Even if S251112cm is found to be terrestrial in origin, we must still consider such signals, as it is likely that more S251112cm-like signals will be reported, especially given planned upgrades to LIGO (e.g., A-sharp; \cite{gupta+24}) and planned future detectors such as ET and CE. Given the importance of identifying a SSM CO---and the substantial physical uncertainties such a discovery would introduce---it is essential that candidate binaries be analyzed with great care to ensure a confident detection and to rule out possibilities with known astrophysical sources.

\begin{figure}
    \centering
    \includegraphics[width=\linewidth]{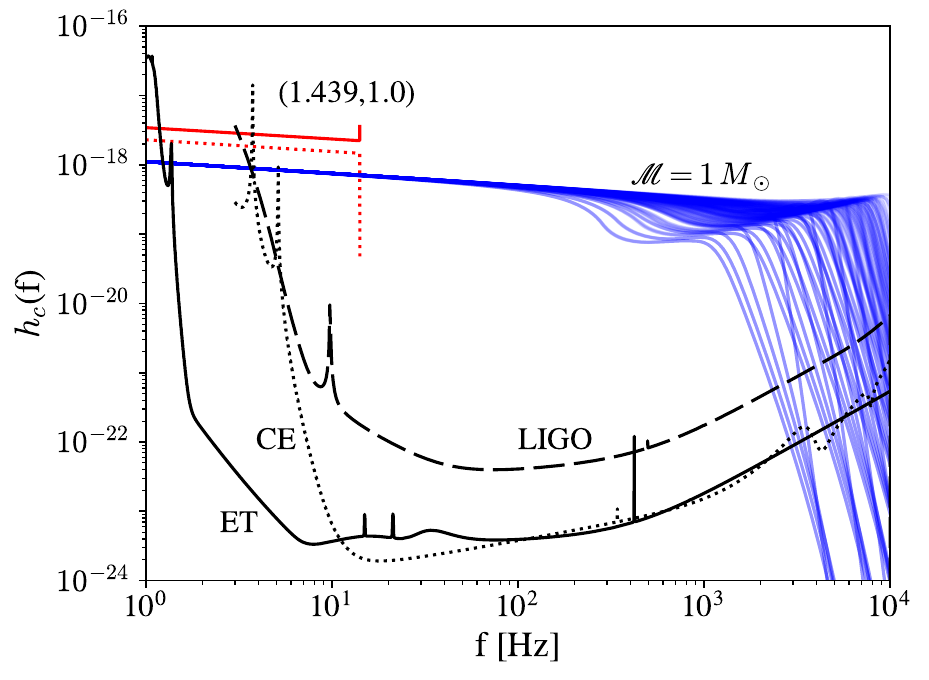}
    \caption{Characteristic strain curve for the physical system (e.g., excluding SSM COs) with the highest cutoff frequency, $f_{\rm max},$ compared to terrestrial detector sensitivity curves and template sensitivity curves from $\Mc=1\,M_\odot$ systems with varying mass ratio and spin configurations. The solid red line is the waveform assuming Jeans Mode while the dotted line assumes Isotropic Re-emission. $\Mc=1\,M_\odot$ waveform templates are calculated using \cite{pratten+20}.}
    \label{fig:Fig_7}
\end{figure}

The highest-frequency gravitational wave produced from WD-NS/BH in our tests reaches $f_{\rm max}\approx15\unit{Hz}$ (Fig.~\ref{fig:Fig_7}), while CO binaries with chirp masses $\Mc<1\,M_\odot$ are expected to reach frequencies of $>10^3\unit{Hz}$. The system in Figure~\ref{fig:Fig_7} has $\Mwd=1.439\,M_\odot$, $\approx99.9\%$ of our assumed Chandrasekhar mass. Even the heaviest WD we test fails to reach the maximum sensitivity region of LIGO, let alone the frequencies characteristic of SSM systems. 

Even if a terrestrial detector were to detect a binary as in Figure~\ref{fig:Fig_7}, the Jeans Mode system has sub-solar effective chirp mass \cite{tauris_18} for less than one second before WD engulfment; effective chirp mass is never sub-solar for the Isotropic Re-emission system. Meanwhile, binaries with truly sub-solar WDs $(\Mwd<1\,M_\odot)$ have $f_{\rm max}$ at least an order of magnitude too low to be detected by LIGO. 

We conclude that WD-NS/BH systems are unable to masquerade as SSM CO gravitational wave events in terrestrial detectors.

\section{Conclusion}\label{sec:conclusion}
In this paper, we examine gravitational waveforms for circular white-dwarf--compact-object binaries undergoing mass transfer. We develop a novel semi-analytic code integrating mass transfer dynamics with gravitational wave radiation reaction at the 2.5 Post-Newtonian order to evolve such systems and calculate their gravitational waveforms (\S\ref{sec:methods}). We use results from these waveforms to characterize binary evolution and to determine observational feasibility.

White-dwarf--compact-object binaries demonstrate distinct, archetypal behavior in their evolution and gravitational waveforms. The binary slowly hardens under radiation reaction until reaching a ``turnaround" frequency, $f_{\rm max}$. In $h_c$ tracks, this manifests as an $h_c\propto f^{-1/6}$ inspiral with a sharp peak and maximum frequency cutoff at $f_{\rm max}$. The system then experiences an ``outspiral" phase during which the binary quickly widens, and the white dwarf eventually undergoes tidal disruption or loses degeneracy pressure. The waveform of this outspiral is characterized by an orders-of-magnitude drop in $h_c$ for frequencies just smaller than $f_{\rm max}$, a leveling off, and finally a truncation in $h_c$ at low frequencies when the white dwarf ceases to exist (\S\ref{sec:results-gw-char}).

The time between mass transfer beginning and the end of the system varies by orders of magnitude between the lowest- and highest-mass white dwarfs, with low-$\Mwd$ systems living much longer than high-$\Mwd$ systems (\S\ref{sec:results-evolution}). Frequency also varies by orders of magnitude with $f_{\rm max}$ for the lowest mass white dwarfs near LISA peak sensitivity and ultra-relativistic white dwarfs approaching the lower edge of the terrestrial detector band. The mass of the compact object partner contributes to determining gravitational wave strength, but is generally only an order-of-magnitude effect (\S\ref{sec:results-gw-trends}).

LISA will have a horizon distance of up to $\sim30\unit{Mpc}$ for relativistic white-dwarf--high-mass-black-hole binaries. However, white-dwarf--neutron-star/black-hole end stage events are rare enough such that this horizon is insufficient for LISA to observe such an event during its lifetime. Detection by terrestrial observatories is even more unlikely as their horizon for detection is sub-Galactic. However, LGWA and DECIGO are primed to detect such events. LGWA will observe tens of white-dwarf--neutron-star end-stage inspirals annually. In DECIGO, white-dwarf--compact-object inspirals will be so prevalent that improvements in data processing will be necessary to analyze them. Weaker outspiral phases will also be detectable in DECIGO with thousands of white-dwarf--neutron-star and several white-dwarf--black-hole outspirals observed per year (\S\ref{sec:discussion-rates}).

A diverse array of ultimate fates is possible for white-dwarf--compact-object systems including true mergers, WD engulfment, tidal disruption events, and ultracompact X-ray binaries. Initial system conditions and angular-momentum-loss mode determine system fates. However, angular-momentum-loss mechanisms remain highly uncertain for white-dwarf--compact-object systems undergoing super-Eddington mass transfer. Further modeling is necessary to robustly constrain end stage fates (\S\ref{sec:discussion-L}).

The highest frequency gravitational waveforms tested have $f_{\rm max}\approx 15\unit{Hz}$ and are produced by a white dwarf with mass $\sim 99.9\%$ of the Chandrasekhar limit. Even these systems do not reach the maximum sensitivity band for terrestrial detectors. White-dwarf--compact-object binaries are not effective mimics for sub-solar-chirp-mass binaries in the LVK band (\S\ref{sec:discussion-SSM}).

\section*{Acknowledgments}
The authors would like to thank the anonymous referee for constructive comments leading to increased clarity in this manuscript. TSW would like to thank Michael Eracleous, Sarah Shandera, James DeLaunay, and Eric Ford for valuable discussions and feedback leading to the development of this manuscript.

TSW and DJ were supported by NSF grants AST-2307026 and AST-2407298 at the Pennsylvania State University (PSU).
DJ was also supported by a KIAS Individual Grant, PG088301.
DR acknowledges support from the U.S.~Department of Energy, Office of Science, Division of Nuclear Physics under Award Number(s) DE-SC0024388, and from the National Science Foundation under Grants PHY-2020275, PHY-2116686, PHY-2407681, PHY-2512802, and PHY-2621752.

Parts of the code were developed using computational resources provided by Astronomy 528 \cite{ford_25} at the Pennsylvania State University. The authors recognize the Penn State Institute for Computational and Data Sciences (ICDS)  (RRID:SCR\_025154) for providing access to computational research infrastructure (RRID:SCR\_026424).

Our code is publicly available at \cite{weaver+26_code}.

\appendix
\section{Stopping Conditions}
\label{sec:app-stopping}
There are two physical scenarios under which our integration ceases: contact (i.e., a merger between the WD and CO or WD engulfment of the CO) or a change in WD equation of state. We consider the system to have merged or have undergone engulfment if the CO's ISCO radius (see Eqn.~\ref{eqn:ISCO}) or the WD radius (see Eqn.~\ref{eqn:Rwd0}), respectively, exceeds the binary separation (\S\ref{sec:discussion-L-merger}). In these cases, Equation~\ref{eqn:da} is no longer a good approximation due to either a strong-gravity-induced plunge \cite{misner+73,ori+00} or common-envelope physics \cite{ivanova+13, ropke+23}. In practice, for COs of stellar BH mass or below, the ISCO radius is not reached; IMBH-scale masses are required for WDs to reach ISCO (Fig.~\ref{fig:Fig_8}).

In the case that binary separation increases and a merger is averted, the integration ceases when the WD loses enough mass such that its equation of state is no longer dominated by degeneracy pressure but instead by ideal gas pressure. Then, we say that the remnant is no longer a WD---instead something more akin to an extremely hot gas giant or brown dwarf---and Equation~\ref{eqn:Rwd0} no longer applies to it. The ideal gas pressure and non-relativistic degeneracy pressure are, respectively \cite{prialnik_09}:
\begin{align}
    P_{\rm IG}&=\frac{k_B\rho T}{\mu m_p}\label{eqn:IG}\\
    P_{\rm deg.}&=\frac{h^2}{20m_e}\parfrac{3}{\pi}^{2/3}\parfrac{\rho}{\mu m_p}^{5/3}\label{eqn:nr}
\end{align}
where $m_{e\,(p)}$ is electron (proton) mass, $k_B$ is the Boltzmann constant, $h$ is the Planck constant, and $\frac{1}{\mu}=\frac{1}{\mu_I}+\frac{1}{\mu_e}$. In reality, there will be regions of a WD where different equations of state dominate---in standard WDs, the outermost layer forms a thin, low-density atmosphere dominated by ideal gas pressure \cite{rohrman_01, bedard_24}. As a zeroth-order approximation, we use the average density of the WD implied by Equation~\ref{eqn:Rwd0}:
\begin{equation}
    \langle\rho\rangle_{\rm WD}\approx\frac{3}{4\pi \pars{7792\unit{km}}^3}\frac{\Mwd^2}{0.7\,M_\odot}\bracs{1-\parfrac{\Mwd}{M_{\rm Ch}}^{4/3}}\parfrac{\mu_e}{2}^{-1}\label{eqn:rho_WD}
\end{equation}
to determine the boundary between a body dominated by degenerate and ideal gas equations of state. By combining Equation~\ref{eqn:rho_WD} with~\ref{eqn:IG} and~\ref{eqn:nr} and assuming a carbon-oxygen WD ($\mu=\frac{7}{4}$), we find that the minimum WD mass is:
\begin{equation}
    M_{\rm WD,\,min.}\approx9\sn{-3}\,M_\odot\parfrac{T_c}{10^6\unit{K}}^{3/4}\label{eqn:M_WD_min}
\end{equation}
where $T_c$ is the temperature of the WD's nearly isothermal core \cite{bhatt+18}. WDs in WD-CO binaries formed via stellar evolution (especially WD-NS binaries) are likely to be young and, therefore, hot ($\sim10^7\unit{K}$; \cite{panei+00, chen+11}). Thus, we take $M_{\rm WD,\,min.} =0.05\,M_\odot$ for our minimum mass cutoff.

\begin{figure}
    \centering
    \includegraphics[width=\linewidth]{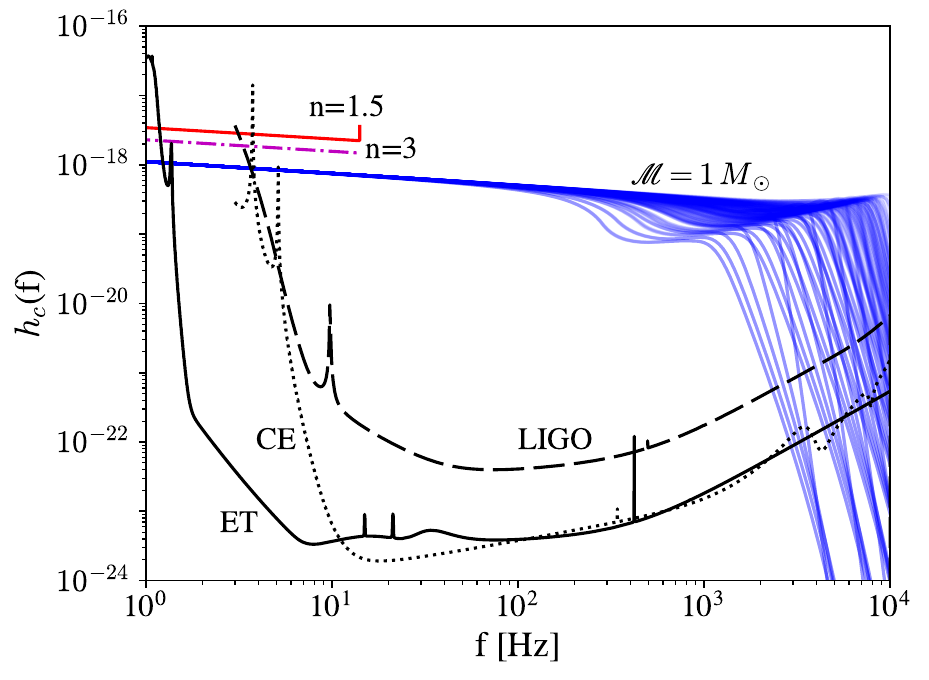}
    \caption{Same as Figure~\ref{fig:Fig_7} but includes an $n=3$ polytrope model used for the WD (violet dash-dotted line). Because the $n=3$ WD is more centrally concentrated, the effects of mass transfer are smaller than in the $n=1.5$ case. The WD engulfs its NS partner before deviation from point-mass inspiral kinematics becomes evident.}
    \label{fig:Fig_10}
\end{figure}

\section{Effect of Mass Transfer Rate}
\label{sec:app-MT}
In this section, we briefly estimate the effect of the choice of mass transfer coefficient $A=10$ on the turnaround frequency, $f_{\rm max}$. In the case with $\dot a_{\rm MT}>0$ (e.g., Jeans Mode; see Eqn.~\ref{eqn:da_MT}), $f_{\rm max}$ is reached when the term in curly brackets of Equation~\ref{eqn:da} is zero. Furthermore, from Figure~\ref{fig:Fig_1}, we find that key parameters from Equation~\ref{eqn:da} are nearly constant during the $\dot a<0$ inspiral phase. Let $\Delta a$ be the hardening in separation from initial Roche Lobe Overflow (i.e., $a=a_{\rm RLOF}-\Delta a$), and let $\delta R$ be the depth below the WD's surface. Near the surface, density goes as $\rho_{\rm WD}\pars{R_{\rm WD}-\delta R}\propto\pars{\delta R}^{n}$ for polytropic index $n$ Equation~\ref{eqn:dMwd} becomes:
\begin{equation}
    |\dot M_{\rm WD}|\propto A\pars{\Delta a}^{1+n}\label{eqn:dM_prop}
\end{equation}
where the proportionality is nearly constant during the inspiral. Then, solving $\dot a=0$ for constant values in Equation~\ref{eqn:da} yields:
\begin{equation}
    \pars{\Delta a}_A \propto A^{-1/(1+n)}\label{eqn:a_A}
\end{equation}
where here we have let $\pars{\Delta a}_A$ mean the maximum value of $\Delta a$ for an implicit configuration of $(\Mwd,\,\Mns)$ and a fixed value of $A$.

Letting $\Delta f=f_{\rm max}-f_{\rm RLOF}$, we have that:
\begin{equation}
    \parfrac{\Delta f}{f}=-\frac{3}{2}\parfrac{\Delta a}{a}\label{eqn:diff_kepler}
\end{equation}
Finally, letting $\parfrac{\Delta f}{f}_{10}:=-\frac{3}{2}\parfrac{\Delta a}{a}_{10}$, Equations~\ref{eqn:a_A} and~\ref{eqn:diff_kepler} imply:
\begin{equation}
    \parfrac{\Delta f}{f}=\parfrac{\Delta f}{f}_{10}\parfrac{A}{10}^{-1/(1+n)}\label{eqn:diff_kepler_A}
\end{equation}
This is to say that taking $A=10\pm 0.5\unit{dex}$ as an order-of-magnitude estimate, $\parfrac{\Delta f}{f}$ varies by $ 0.4\unit{dex}$ between low and high estimates for $A$ with $n=1.5$. 

From our integrations, we have that $\parfrac{\Delta f}{f}_{10}\approx1$--$5\%$, meaning that choice of $A$ contributes a systematic error of $\delta\parfrac{\Delta f}{f}\approx 0.006$--0.03 in determining maximum frequency where we have taken $A\sim10$ as an order-of-magnitude value.

\section{Choice of Polytropic Index}
\label{sec:app-PI}
While non-relativistic, low-mass WDs obey an $n=1.5$ polytropic density profile, the relativistic degenerate equation of state of high-mass WDs imposes an index of $n=3$ \cite{chandrasekhar_31, shapiro+83}. 

Between the non-relativistic and ultra-relativistic limits, the density profile of a WD is generally determined by integrating the Chandrasekhar equation of state \cite{chandrasekhar_35} (see e.g., \cite{boshkayev+15,boshkayev+17,zeyauddin+2026} for modeling applications). This transition region is important for the case of initially ultra-relativistic WDs which lose mass and become non-relativistic. However, to model this transition phase fully consistently requires the calculation of many intermediate models. This level of modeling is beyond the scope of this paper.

Instead, we could choose to model WDs above a certain mass threshold with $n=3$ and below the threshold as $n=1.5$. However, Equation~\ref{eqn:da} requires differentiability; a sudden change between $n=3$ and $n=1.5$ profiles results in numerical instability within our model.

Our model adopts an $n=1.5$ profile consistently across all tests for simplicity. Uniformly adopting a single index allows us to bypass the modeling complications of a changing profile due to mass loss.

Our key rate forecast results are unaffected by the choice of polytropic index in high-mass WDs at the precision reported in \S\ref{sec:discussion-rates}, and the effects of angular momentum loss mode on system outcomes explored in \S\ref{sec:discussion-L} are much greater than those of polytropic index.

The context in which polytropic index is most important in our study is \S\ref{sec:discussion-SSM} which explicitly discusses the outcome of a system with ultra-relativistic WDs. However, as shown in Figure~\ref{fig:Fig_10}, even when modeled with physically consistent $n=3$, the key result of the analysis is unchanged; a WD-CO binary cannot masquerade as a SSM CO binary.

\bibliographystyle{apsrev4-2}
\bibliography{references}

\end{document}